\documentclass[12pt]{article}
\newcommand{\sect}[1]{\setcounter{equation}{0}\section{#1}}

\def\[{{[}}

\def\CP{{\cal P}}

\def\top{\hspace{-6mm}}

\def\1{\mathbb I}
\newcommand{\bea}{\begin{eqnarray}}
\newcommand{\eea}{\end{eqnarray}}

\renewcommand{\CP}[1]{\mathbb{C}P^#1}
\newcommand{\bbR}{{\mathbb R}}
\newcommand{\bbZ}{{\mathbb Z}}

\renewcommand\top{\hspace{-6mm}}

\def\fnote#1#2{\begingroup\def\thefootnote{#1}\footnote{#2}\addtocounter
{footnote}{-1}\endgroup}

\usepackage{amsbsy,amssymb,latexsym}

\begin{document}
\begin{flushright}
OCU-PHYS 207 \\
hep-th/0402199\\

\end{flushright}
\vspace{20mm}

\begin{center}
{\bf\Large
New Infinite Series of
Einstein Metrics

on
Sphere Bundles
from
AdS Black Holes
}

\vspace{20mm}
Yoshitake Hashimoto\fnote{$\dagger$}{
\texttt{hashimot@sci.osaka-cu.ac.jp}
},
Makoto Sakaguchi\fnote{$\star$}{
\texttt{msakaguc@sci.osaka-cu.ac.jp}
}
and
Yukinori Yasui\fnote{$\ast$}{
\texttt{yasui@sci.osaka-cu.ac.jp}
}
\vspace{10mm}

\textit{
${{}^\dagger}$
Department of Mathematics, Osaka City University
}
\vspace{2mm}

\textit{
${{}^\star}$
Osaka City University
Advanced Mathematical Institute (OCAMI)
}
\vspace{2mm}

${}^\ast$
\textit{
Department of Physics, Osaka City University
}

\vspace{5mm}

\textit{
Sumiyoshi,
Osaka 558-8585, JAPAN}
\end{center}
\vspace{25mm}

\begin{abstract}
A new infinite series of
Einstein metrics is
 constructed explicitly on $S^2\times S^3$,
 and the non-trivial $S^3$-bundle over $S^2$,
containing infinite numbers of inhomogeneous ones.
They appear as a certain limit of a nearly extreme 
5-dimensional AdS Kerr black hole.
In the special case,
the metrics reduce to the homogeneous Einstein metrics
studied by Wang and Ziller.
We also construct an inhomogeneous Einstein metric
on the non-trivial $S^{d-2}$-bundle over $S^2$
from a $d$-dimensional AdS Kerr black hole.
Our construction is a higher dimensional version
of the method of
Page,
which gave an inhomogeneous Einstein metric
on
$\CP2\sharp\overline{\CP2}$.

\end{abstract}
\newpage

\sect{Introduction}
Anti-de Sitter (AdS) spaces have attracted renewed interests after
the AdS/CFT correspondence conjecture \cite{Maldacena},
which relates
the properties of the supergravity on AdS
and those of 
the strongly coupled gauge theory
on the AdS boundary.
For example,
the Hawking-Page phase transition \cite{HP}
between AdS and AdS Schwarzschild black holes
was interpreted \cite{Witten} as
the phase transition between  confining and deconfining
phases of the dual gauge theory.
Motivated by this,
the study of AdS black holes has been extended in various directions.
Among them,
AdS Kerr black holes
with two angular momenta in 5-dimensions,
as well as ones with one angular momentum in $d$-dimensions were
constructed in \cite{Hawking}.

On the other hand, 
an inhomogeneous Einstein metric on $\CP2\sharp\overline{\CP2}$
was constructed by Page \cite{Page}.
This metric is of cohomogeneity one with principal orbits $S^3$.
It was obtained as a certain limit of the 4-dimensional de Sitter
black hole together with the Wick rotation.
It should be emphasized that this metric is the first example of
inhomogeneous Einstein metrics.
Furthermore,
B\"ohm proved the existence of an infinite series of Einstein metrics
of cohomogeneity one with positive scalar curvature
on $S^N$ ($5\le N\le 9$)
and $S^{N_1+1}\times S^{N_2}$ 
($5\le N_1+N_2+1\le 9$, $N_1>1$, $N_2>1$) \cite{Bohm}.

Combining these two observations,
we explicitly construct
new Einstein metrics
with positive scalar curvature
on sphere bundles,
applying the method developed in \cite{Page}
to the 5-dimensional  AdS Kerr black hole
with two angular momenta
and the $d$-dimensional AdS Kerr black hole
with one angular momentum
constructed in \cite{Hawking}.
In summary, we will construct
\begin{itemize}
  \item an infinite series of  Einstein metrics
on $S^2\times S^3$ and $S^2\widetilde{\times} S^3$
(the non-trivial $S^3$-bundle over $S^2$)
parameterized by a pair of integers $(k_1,k_2)$.
The bundle type depends on the parity
of $k_1+k_2$;
It is trivial if $k_1+k_2$ is even,
and is non-trivial otherwise.
When $k_1\neq k_2$, the metrics are inhomogeneous (see Theorem 1).
When $k_1= k_2$, they are homogeneous Einstein metrics on $S^2\times S^3$
(see Theorem 2).

  \item an inhomogeneous Einstein
metric on $S^2\widetilde{\times} S^{d-2}$,
the non-trivial $S^{d-2}$-bundle over $S^2$ (see Theorem 3).

\end{itemize}

It should be noticed that
the metrics on
$S^2\widetilde{\times} S^3$
in Theorem 1
obviously are not included in the case of
B\"ohm's existence theorem
and that
the metrics on $S^2\times S^3$ are apparently different from
the ones which are proved to exist in \cite{Bohm}.
The metrics in Theorem 2 coincide with the homogeneous Einstein
metrics on $M_{k,1}^{1,1}$,
circle bundles over $\CP1\times\CP1$
studied by Wang and Ziller \cite{Wang}.
The metric in Theorem 3
is of cohomogeneity one with
principal orbits $S^3\times S^{d-4}$ if $d\ge 5$.
In the case of $d=4$,
it reproduces the Page metric on 
$S^2\widetilde{\times}S^2=\CP2\sharp\overline{\CP2}$
with principal orbits $S^3$.

This paper is organized as follows.
In section 2, we apply the method \cite{Page}
to the 5-dimensional AdS Kerr black hole with two angular momenta,
and obtain the first class of the Einstein metrics.
We will derive the second class of the Einstein metrics
from the $d$-dimensional AdS Kerr black hole \cite{Hawking}
with one angular momentum in section 3.

\sect{5-dimensional Einstein Metrics}
The 5-dimensional Euclidean de Sitter (dS) Kerr metric may be
extracted from the Lorentzian AdS Kerr metric \cite{Hawking}
by the substitution
$
t\mapsto -i\tau,~~~
a\mapsto -i\alpha,~~~
b\mapsto -i\beta
$
with $l\mapsto -il$ (AdS $\mapsto$ dS),
\begin{eqnarray}
g_5&=&
\frac{\Delta_r}{\rho^2}\biggl[
 d\tau
 -\frac{\alpha\sin^2\theta}{\Xi_\alpha}d\phi
 -\frac{\beta\cos^2\theta}{\Xi_\beta}d\psi
 \biggr]^2
+\frac{\Delta_\theta\sin^2\theta}{\rho^2}\biggl[
 \alpha d\tau
 +\frac{r^2-\alpha^2}{\Xi_\alpha}d\phi
 \biggr]^2
\nonumber\\&&
+\frac{\Delta_\theta\cos^2\theta}{\rho^2}
 \biggl[
 \beta d\tau
 +\frac{r^2-\beta^2}{\Xi_\beta}d\psi
 \biggr]^2
\nonumber\\&&
-\frac{1-r^2l^2}{r^2\rho^2}
 \biggl[
 \alpha\beta d\tau
 +\frac{\beta(r^2-\alpha^2)\sin^2\theta}{\Xi_\alpha}d\phi
 +\frac{\alpha(r^2-\beta^2)\cos^2\theta}{\Xi_\beta}d\psi
 \biggr]^2
\nonumber\\&&
+\frac{\rho^2}{\Delta_r}dr^2
+\frac{\rho^2}{\Delta_\theta}d\theta^2,
\label{bh}
\end{eqnarray}
where
\begin{eqnarray}
\rho^2&=&
 r^2
 -\alpha^2\cos^2\theta
 -\beta^2\sin^2\theta,
\nonumber\\
\Delta_r&=&
 \frac{1}{r^2}(r^2-\alpha^2)(r^2-\beta^2)(1-r^2l^2)-2M,
\nonumber\\
\Delta_\theta&=&
 1
 -\alpha^2l^2\cos^2\theta
 -\beta^2l^2\sin^2\theta
\end{eqnarray}
with
parameters
$\Xi_\alpha=1-\alpha^2l^2$
and
$\Xi_\beta=1-\beta^2l^2$.
The radii of the horizons are given by the roots of $\Delta_r=0$.
If there is a double root $r_0$,
the black hole becomes an extreme black hole,
in which the number of the free parameters reduces.
The parameters
$r_0,~M,~\Xi_\alpha$ and $\Xi_\beta$
are written in terms
of $l$ and the dimensionless parameters
\begin{eqnarray}
\nu_1=\alpha/r_0,~~~
\nu_2=\beta/r_0,
\end{eqnarray}
as\fnote{$\ddagger$}{
The suffix $0$ is added for extreme parameters.
}
\begin{eqnarray}
r_0&=&\biggl[
 \frac{1-\nu_1^2\nu_2^2}{2-\nu_1^2-\nu_2^2}
 \biggr]^{1/2}l^{-1},
\nonumber\\
M_0&=&
 \frac{(1-\nu_1^2)^2(1-\nu_2^2)^2(1-\nu_1^2\nu_2^2)}
 {2(2-\nu_1^2-\nu_2^2)^2}
 l^{-2},
\nonumber\\
\Xi_\alpha^0&=&
 1-\frac{\nu_1^2(1-\nu_1^2\nu_2^2)}{2-\nu_1^2-\nu_2^2},
\nonumber\\
\Xi_\beta^0&=&
 1-\frac{\nu_2^2(1-\nu_1^2\nu_2^2)}{2-\nu_1^2-\nu_2^2}.
\end{eqnarray}
In terms of these parameters,
we have
\begin{eqnarray}
\Delta_r&=&
 -(r-r_0)^2\tilde{\Delta}(r),
\nonumber\\
\tilde\Delta(r)&=&
 \frac{1}{r^2}(r+r_0)^2
 (
l^2r^2
-\nu_1^2\nu_2^2
).
\end{eqnarray}

Let us consider a nearly extreme black hole;
it has two horizons located at $r=r_1\equiv r_0-\varepsilon$ and 
$r=r_2\equiv r_0+\varepsilon$,
where the parameter $\varepsilon$ represents a small deviation
from the extreme
black hole.
In the region between two horizons,
$\Omega(\varepsilon)=\{r|r_1\le r \le r_2\}$,
we introduce
a new radial coordinate $\chi$ ($0\le \chi\le \pi$) by
\begin{eqnarray}
r=r_0-\varepsilon\cos\chi.
\end{eqnarray}
The $\Delta_r=-(r-r_1)(r-r_2)\tilde\Delta(r)$
restricted to $\Omega(\varepsilon)$
is then given by
\begin{eqnarray}
\Delta_r=\varepsilon^2\tilde\Delta(r_0)\sin^2\chi+O(\varepsilon^3),
\end{eqnarray}
where
\begin{eqnarray}
\tilde\Delta(r_0)=
\frac{4(1+\nu_1^2\nu_2^4+\nu_1^4\nu_2^2-3\nu_1^2\nu_2^2
)}{2-\nu_1^2-\nu_2^2}.
\end{eqnarray}
We consider the limit
$\varepsilon\to 0$
of the metric (\ref{bh}) in the region $\Omega(\varepsilon)$.
In this limit, the term proportional to $\Delta_r$
vanishes and so leads to a singularity of the metric.
To avoid this, 
we define a rescaled time coordinate
\begin{eqnarray}
\eta=
\frac{\varepsilon\tilde\Delta(r_0)}{r_0^2(1-\nu_1^2)(1-\nu_2^2)}\tau
+O(\varepsilon^2).
\end{eqnarray}
It is also convenient to define the azimuthal angles
\begin{eqnarray}
\phi_1=\phi+\frac{\alpha\Xi_\alpha}{r_1^2-\alpha^2}\tau,~~~
\phi_2=\psi+\frac{\beta\Xi_\beta}{r_1^2-\beta^2}\tau.
\end{eqnarray}
The new coordinates $(\eta, \chi, \phi_1,\phi_2,\theta)$
are then well-behaved local coordinates
in the limit $\varepsilon \to 0$:
\begin{eqnarray}
(a)&&
\frac{\Delta_r}{\rho^2}\biggl[
 d\tau
 -\frac{\alpha\sin^2\theta}{\Xi_\alpha}d\phi
 -\frac{\beta\cos^2\theta}{\Xi_\beta}d\psi
 \biggr]^2
\to 
\frac{\rho_0^2}{\tilde\Delta(r_0)}\sin^2\chi d\eta^2
\\
(b)&&
\frac{\Delta_\theta\sin^2\theta}{\rho^2}\biggl[
 \alpha d\tau
 +\frac{r^2-\alpha^2}{\Xi_\alpha}d\phi
 \biggr]^2
\nonumber\\&&
\to
\frac{r_0^4(1-\nu_1^2)^2}{\rho_0^2}
\Delta_\theta^0\sin^2\theta
\biggl[
\frac{d\phi_1}{\Xi_\alpha^0}
-\frac{4\nu_1(1-\nu_2^2)}{(1-\nu_1^2)\tilde\Delta(r_0)}
\sin^2\frac{\chi}{2} d\eta
\biggr]^2
\\
(c)&&
\frac{\Delta_\theta\cos^2\theta}{\rho^2}
 \biggl[
 \beta d\tau
 +\frac{r^2-\beta^2}{\Xi_\beta}d\psi
 \biggr]^2
\nonumber\\&&
\to
\frac{r_0^4(1-\nu_2^2)^2}{\rho_0^2}
\Delta_\theta^0\cos^2\theta
\biggl[
\frac{d\phi_2}{\Xi_\beta^0}
-\frac{4\nu_2(1-\nu_1^2)}{(1-\nu_2^2)\tilde\Delta(r_0)}
\sin^2\frac{\chi}{2} d\eta
\biggr]^2
\\
(d)&&
\frac{1-r^2l^2}{r^2\rho^2}
 \biggl[
 \alpha\beta d\tau
 +\frac{\beta(r^2-\alpha^2)\sin^2\theta}{\Xi_\alpha}d\phi
 +\frac{\alpha(r^2-\beta^2)\cos^2\theta}{\Xi_\beta}d\psi
 \biggr]^2
\nonumber\\&&
\to
\frac{r_0^4(1-\nu_1^2)(1-\nu_2^2)}{\rho_0^2(2-\nu_1^2-\nu_2^2)}
\biggl[
\frac{\nu_2(1-\nu_1^2)}{\Xi_\alpha^0}\sin^2\theta d\phi_1
+\frac{\nu_1(1-\nu_2^2)}{\Xi_\beta^0}\cos^2\theta d\phi_2
\nonumber\\&&\hspace{40mm}
-\frac{4\nu_1\nu_2\rho_0^2}{\tilde\Delta(r_0)r_0^2}
\sin^2\frac{\chi}{2}d\eta
\biggr]^2
\\
(e)&&
\frac{\rho^2}{\Delta_r}dr^2
\to
\frac{\rho^2_0}{\tilde\Delta(r_0)}d\chi^2
\\
(f)&&
\frac{\rho^2}{\Delta_\theta}d\theta^2
\to
\frac{\rho^2_0}{\Delta_\theta^0}d\theta^2,
\end{eqnarray}
where
\begin{eqnarray}
\rho_0^2&=&
 r_0^2(1-\nu_1^2\cos^2\theta-\nu_2^2\sin^2\theta),\\
\Delta_\theta^0&=&
1-\frac{1-\nu_1^2\nu_2^2}{2-\nu_1^2-\nu_2^2}
(\nu_1^2\cos^2\theta+\nu_2^2\sin^2\theta).
\end{eqnarray}

Finally,
we obtain a metric with two parameters $\nu_1$ and $\nu_2$
\begin{eqnarray}
g=h^2(\theta)d\theta^2
 +\sum_{i,j=1}^2a_{ij}(\theta)\omega^i\otimes \omega^j
 +b^2(\theta)g_{S^2}.
 \label{metric}
\end{eqnarray}
Here, $g_{S^2}$ represents the standard metric on $S^2$,
\begin{eqnarray}
g_{S^2}=d\chi^2+\sin^2\chi d\eta^2.
\label{gS}
\end{eqnarray}
The 1-forms $\omega^i$ ($i=1,2$) are defined by
\begin{eqnarray}
\omega^i=d\psi_i+k_i\cos\chi d\eta,
\end{eqnarray}
where
\begin{eqnarray}
k_1&=&
\frac{4\nu_1(1-\nu_2^2)\Xi_\alpha^0}{(1-\nu_1^2)\tilde\Delta(r_0)}
=\frac{\nu_1(1-\nu_2^2)(2-\nu_2^2-\nu_1^2\nu_2^2)}
 {1+\nu_1^4\nu_2^2+\nu_1^2\nu_2^4-3\nu_1^2\nu_2^2},
\nonumber
\\
k_2&=&
\frac{4\nu_2(1-\nu_1^2)\Xi_\beta^0}{(1-\nu_2^2)\tilde\Delta(r_0)}
=\frac{\nu_2(1-\nu_1^2)(2-\nu_1^2-\nu_1^2\nu_2^2)}
 {1+\nu_1^4\nu_2^2+\nu_1^2\nu_2^4-3\nu_1^2\nu_2^2}, 
\label{k} 
\end{eqnarray}
and $\psi_i$ are introduced as
\begin{eqnarray}
\phi_i=\frac{1}{2}(\psi_i+k_i\eta).
\end{eqnarray}
The metric components are found to be\fnote{$\S$}{
The metric is symmetric with respect to the exchange
$(\psi_1,\nu_1)\leftrightarrow (\psi_2,\nu_2)$.
Taking account of this symmetry, we often discuss only the one side.
}
\begin{eqnarray}
h^2
&=&
 \frac{1-\nu_1^2\cos^2\theta-\nu_2^2\sin^2\theta}{\Delta_\theta^0},
\label{h^2}\\
a_{11}
&=&
 \biggl(
 \frac{1-\nu_1^2}{2\Xi_\alpha^0}
 \biggr)^2
 \frac{\sin^2\theta}{1-\nu_1^2\cos^2\theta-\nu_2^2\sin^2\theta}
 \biggl[
 \Delta_\theta^0
 -\frac{\nu_2^2(1-\nu_1^2)(1-\nu_2^2)}{2-\nu_1^2-\nu_2^2}\sin^2\theta
 \biggr],
\label{a11}\\
a_{22}
&=&
 \biggl(
 \frac{1-\nu_2^2}{2\Xi_\beta^0}
 \biggr)^2
 \frac{\cos^2\theta}{1-\nu_1^2\cos^2\theta-\nu_2^2\sin^2\theta}
 \biggl[
 \Delta_\theta^0
 -\frac{\nu_1^2(1-\nu_1^2)(1-\nu_2^2)}{2-\nu_1^2-\nu_2^2}\cos^2\theta
 \biggr],
\\
a_{12}
&=&
 -\biggl(
 \frac{\nu_1\nu_2(1-\nu_1^2)^2(1-\nu_2^2)^2}
 {4\Xi_\alpha^0\Xi_\beta^0(2-\nu_1^2-\nu_2^2)}
 \biggr)
 \frac{\sin^2\theta\cos^2\theta}
 {1-\nu_1^2\cos^2\theta-\nu_2^2\sin^2\theta},
\label{a12}\\
b^2
&=&
\frac{1-\nu_1^2\cos^2\theta-\nu_2^2\sin^2\theta}{\tilde\Delta(r_0)}.
\label{b}
\end{eqnarray}

It is straightforward to calculate the Ricci curvature.
We find that
(\ref{metric})
is the Einstein metric
with the
scalar curvature $20(1-\nu_1^2\nu_2^2)/(2-\nu_1^2-\nu_2^2)$.
In the following, we shall consider three cases
for two parameters $\nu_1$ and $\nu_2$;
\begin{quote}
\begin{description}
  \item[Case A]~ $\nu_1^2, \nu_2^2>1 ~\mbox{and}~ \nu_1^2\neq\nu_2^2$,
  \item[Case B]~ $0\le \nu_1^2,\nu_2^2\le 1 ~\mbox{and}~ \nu_1^2\neq \nu_2^2$,
  \item[Case C]~ $\nu_1^2=\nu_2^2\equiv\nu$.
\end{description}
\end{quote}
These conditions ensure that
singularities of the Riemannian curvature disappear,
and further
 the metric components
are non-negative, i.e.,
the eigenvalues of $a_{ij}$ are non-negative and $h^2, b^2> 0$.

Next, we consider the condition
to avoid orbifold singularities,
which restricts the range of the angles $(\theta,\psi_i)$
and the parameters $k_i$.
We calculate the determinant
\begin{eqnarray}
\det(a_{ij})=
\left(
\frac{(1-\nu_1^2)(1-\nu_2^2)}{4\Xi_\alpha^0\Xi_\beta^0}
\right)^2
\frac{\sin^2\theta\cos^2\theta\Delta_\theta^0}{1-\nu_1^2\cos^2\theta-\nu_2^2\sin^2\theta}.
\end{eqnarray}
Now clearly $\theta=0$, $\pi/2$ are zeros of the determinant
and so the range of $\theta$
must be restricted to $0\le \theta\le \pi/2$.
The singularities at $\theta=0$, $\pi/2$ are removable
bolt singularities.
Indeed, near the boundaries the metric behaves
as
\begin{eqnarray}
g&\to&
\frac{1-\nu_1^2}{\Xi_\alpha^0}(
d\theta^2+\frac{1}{4}\theta^2d\psi_1^2
)
+g_{L_{k_2}}
~~~\mbox{for}~~\theta\to 0,
\end{eqnarray}
and
\begin{eqnarray}
g&\to&
\frac{1-\nu_2^2}{\Xi_\beta^0}\biggl(
d(\frac{\pi}{2}-\theta)^2+\frac{1}{4}(\frac{\pi}{2}-\theta)^2d\psi_2^2
\biggr)
+g_{L_{k_1}}
~~~\mbox{for}~~\theta\to \frac{\pi}{2},
\end{eqnarray}
where
\begin{eqnarray}
g_{L_{k_1}}&=&
a_{11}(\pi/2)(d\psi_1+k_1\cos\chi d\eta)^2
+b^2(\pi/2)g_{S^2},\\
g_{L_{k_2}}&=&
a_{22}(0)(d\psi_2+k_2\cos\chi d\eta)^2
+b^2(0)g_{S^2}.
\end{eqnarray}
Thus these singularities are removable
provided that the ranges of $\psi_i$ ($i=1,2$) are chosen
to be
$0\le\psi_i\le4\pi$.
In this range, $(\theta,\psi_1/2)$ as $\theta\to 0$
and $(\pi/2-\theta,\psi_2/2)$ as $\theta\to \pi/2$
are the usual polar coordinates on $\bbR^2$.
We also demand that $k_i$ is integral (see eqn.(\ref{k})),
then the 1-forms $\omega_i=d\psi_i+k_i\cos\chi d\eta$
are identified with connections on the lens spaces
$L_{k_i}=L(k_i,1)=S^3/{\bbZ}_{k_i}$,
and each $k_i$ represents the first Chern number
(or the monopole charge) as a circle bundle on $S^2$.
This yields that
the manifolds near the boundaries
are $ \bbR^2\times L_{k_i}$,
which collapses onto $\{\mbox{point}\}\times L_{k}$
at the boundaries.
\medskip

\top
\underline{\textit{Remark 1.}}~~~
There exists a nontrivial root of $\Delta_\theta^0=0$,
\begin{eqnarray}
\theta=\theta_0,~~~
\cos^2\theta_0=
\frac{2-\nu_1^2-2\nu_2^2+\nu_1^2\nu_2^4}{(1-\nu_1^2\nu_2^2)(\nu_1^2-\nu_2^2)}
\end{eqnarray}
other than $\theta=0$ and $\pi/2$.
In order to avoid curvature singularities,
we restrict $\theta_0$ to be in the region $0<\theta_0<\pi/2$
and the parameters $(\nu_1,\nu_2)$ in the following region:
\begin{quote}
\begin{description}
  \item[Case D]~ $\nu_2^2>1,~\nu_2^2(1+\nu_1^2)<2 
~~~\mbox{or}~~~
\nu_2^2<1,~\nu_2^2(1+\nu_1^2)>2$,
  \item[Case E]~ $\nu_1^2>1,~\nu_1^2(1+\nu_2^2)<2 
~~~\mbox{or}~~~
\nu_1^2<1,~\nu_1^2(1+\nu_2^2)>2$.
\end{description}
\end{quote}
Then, the metric is regular if we choose
the range as
$0\le \theta\le\theta_0$ for the case D,
and $\theta_0\le \theta\le \pi/2$ for the case E.
However one can not resolve the orbifold singularity
at the boundary $\theta=\theta_0$
like the cases A, B and C. 
\medskip

Having constructed the Einstein metric locally,
we now proceed to the global issue
for the cases A, B and C.
It is known that 
there are two inequivalent classes of $S^N$-bundles
over $S^2$.
The $S^N$-bundles over $S^2=D_+\cup D_-$ ($D_\pm$ denote hemispheres)
are obtained by attaching $D_+\times S^N$
and $D_-\times S^N$ as
\begin{eqnarray}
&&(x,\xi)\sim (x,\gamma(x)\xi),\nonumber\\
&&x\in D_+\cap D_-=S^1,~~~
\gamma: S^1\mapsto SO(N+1).
\end{eqnarray}
They are classified by $[\gamma]\in \pi_1(SO(N+1))=\bbZ_2$.
Globally our metrics (\ref{metric}) can be regarded as those
on $S^3$-bundles
over $S^2$.

We will discuss the cases A, B and C separately below.
\medskip

\top
\underline{\textbf{Case A}}

We can write the metric (\ref{metric}) with (\ref{gS})-(\ref{b})
as
\begin{eqnarray}
g_{\nu_1\nu_2}=
h^2(\theta)d\theta^2
+\sum_{i,j=1}^{2}a_{ij}(\theta)\omega^i\otimes\omega^j
+b^2(\theta)g_{S^2},
\label{g:nu1nu2}
\end{eqnarray}
where
\begin{eqnarray}
h^2&=&
\frac{1-\nu_1^2\cos^2\theta-\nu_2^2\sin^2\theta}
{1-\mu_1^2\cos^2\theta-\mu_2^2\sin^2\theta},
\\
a_{11}&=&
\frac{1}{4}\left(
\frac{\nu_1(1-\nu_2^2)(2-\nu_1^2-\nu_2^2)}
{1+\nu_1^4\nu_2^2+\nu_1^2\nu_2^4-3\nu_1^2\nu_2^2}
\right)^2
\frac{(1-\mu_1^2\cos^2\theta-\nu_2^2\sin^2\theta)\sin^2\theta}
{1-\nu_1^2\cos^2\theta-\nu_2^2\sin^2\theta},
\\
a_{22}&=&
\frac{1}{4}\left(
\frac{\nu_2(1-\nu_1^2)(2-\nu_1^2-\nu_2^2)}
{1+\nu_1^4\nu_2^2+\nu_1^2\nu_2^4-3\nu_1^2\nu_2^2}
\right)^2
\frac{(1-\nu_1^2\cos^2\theta-\mu_2^2\sin^2\theta)\cos^2\theta}
{1-\nu_1^2\cos^2\theta-\nu_2^2\sin^2\theta},
\\
a_{12}&=&
-\frac{\nu_1^2\nu_2^2(1-\nu_1^2)^2(1-\nu_2^2)^2(2-\nu_1^2-\nu_2^2)}
{4(1+\nu_1^4\nu_2^2+\nu_1^2\nu_2^4-3\nu_1^2\nu_2^2)^2}
\frac{\sin^2\theta\cos^2\theta}{1-\nu_1^2\cos^2\theta-\nu_2^2\sin^2\theta},
\\
b^2&=&
\frac{(2-\nu_1^2-\nu_2^2)(1-\nu_1^2\cos^2\theta-\nu_2^2\sin^2\theta)}
{4(1+\nu_1^4\nu_2^2+\nu_1^2\nu_2^4-3\nu_1^2\nu_2^2)}
\end{eqnarray}
with
\begin{eqnarray}
\mu_1^2=\frac{\nu_1^2(1-\nu_1^2\nu_2^2)}{2-\nu_1^2-\nu_2^2},~~~~
\mu_2^2=\frac{\nu_2^2(1-\nu_1^2\nu_2^2)}{2-\nu_1^2-\nu_2^2}.
\end{eqnarray}
The metric (\ref{g:nu1nu2})
can be regarded as one on the associated $S^3$-bundle
of the principal $T^2$-bundle over $S^2$
with the Euler classes 
$(k_1,k_2)\in H^2(S^2,\bbZ)^{\oplus 2}=\bbZ\oplus\bbZ$.
The invariant $[\gamma]\in \bbZ_2$
of the $S^3$-bundle is given by
\begin{eqnarray}
[\gamma]=k_1+k_2~~\mbox{mod}~2.
\end{eqnarray}
The connection $\omega=\omega^1\oplus\omega^2$
is given by
\begin{eqnarray}
\omega^i=d\psi_i+\cos\chi d\eta,~~~~
0\le\psi_i\le 4\pi/|k_i|,
\end{eqnarray}
where we have rescaled the torus angles as $\psi_i\mapsto k_i\psi_i$.

Summarizing the consideration above, we state

\bigskip

\top
\textbf{Theorem 1.}~~~
\textit{Let $\nu_1$ and $\nu_2$
be real numbers
in the region $\nu_1^2, \nu_2^2 >1$
and $\nu_1^2\neq\nu_2^2$
together with the integral conditions;
\begin{eqnarray}
k_1&=&
\frac{\nu_1(1-\nu_2^2)(2-\nu_2^2-\nu_1^2\nu_2^2)}
{1+\nu_1^4\nu_2^2+\nu_1^2\nu_2^4-3\nu_1^2\nu_2^2},
\label{k1}
\\
k_2&=&
\frac{\nu_2(1-\nu_1^2)(2-\nu_1^2-\nu_1^2\nu_2^2)}
{1+\nu_1^4\nu_2^2+\nu_1^2\nu_2^4-3\nu_1^2\nu_2^2},
\label{k2}
\end{eqnarray}
where $(k_1,k_2)\in \bbZ\oplus\bbZ$.
Then,
$\{g_{\nu_1\nu_2}\}$
gives an infinite series of inhomogeneous Einstein metrics
with positive scalar curvature
$20(1-\nu_1^2\nu_2^2)/(2-\nu_1^2-\nu_2^2)$
on $S^3$-bundles over $S^2$.
If the integer $k_1+k_2$ is even (odd),
then the bundle is trivial (non-trivial).}
\bigskip

\top
\underline{\textit{Remark 2.}}~~~
The region $\nu_1>1,~\nu_2>1$
is mapped diffeomorphically onto the region
$k_1>0,~k_2>0,~k_1+k_2>2$ by (\ref{k1}) and (\ref{k2}).
Hence, there exists the unique pair
$(\nu_1,\nu_2)$
for each $(k_1,k_2)\in S$, 
$S=\{ (k_1,k_2)\in\bbZ\oplus\bbZ~|~k_1\neq\pm k_2, k_1\neq 0, k_2\neq 0\}$.
For example, they are numerically evaluated as
\begin{description}
  \item[(i)] $(k_1,k_2)=(1,2)$, ~~$(\nu_1,\nu_2)=(3.31133,~2.14921)$,
  \item[(ii)] $(k_1,k_2)=(1,3)$, ~~$(\nu_1,\nu_2)=(7.68872,~3.06769)$,
  \item[(iii)] $(k_1,k_2)=(2,3)$, ~~$(\nu_1,\nu_2)=(5.85109, 4.13646)$.
\end{description}
\medskip

\top
\underline{\textit{Remark 3.}}~~~
In the limit $(\nu_1,\nu_2)=(\nu_1,\infty)$,
the metric tends to
\begin{eqnarray}
g_{\nu_1\infty}&=&
d\theta^2
+\frac{\sin^2\theta}{4}(d\psi_1+\cos\chi d\eta)^2
+\frac{\cos^2\theta}{4}d\psi_2^2
+\frac{\sin^2\theta}{4}g_{S^2},
\label{g:nu1oo}
\end{eqnarray}
after the rescaling $\nu_1^2g\to g$.
When we make a modification of the range
of $\psi_1$,
\begin{eqnarray}
0\le\psi_1\le
4\pi/|k_1|
\longmapsto 
0\le\psi_1\le 4\pi,~~~~~~
(k_1=\frac{1+\nu_1^2}{\nu_1})
\end{eqnarray}
then (\ref{g:nu1oo})
represents the standard metric on $S^5$.

\medskip

\top
\underline{\textbf{Case B}}

The parameters $\nu_1$ and $\nu_2$
are restricted to $0\le\nu_1^2, \nu_2^2\le 1$
and $\nu_1^2\neq\nu_2^2$.
By (\ref{k})
we find that $0\le |k_i|\le 2$ ($i=1,2$),
hence there are two possibilities for $(|k_1|,|k_2|)$
under the integral condition\fnote{$\diamond$}{
Remember that the metric is symmetric under
$(\psi_1,\nu_1)\leftrightarrow (\psi_2,\nu_2)$.
}:
(i)~$(|k_1|,|k_2|)=(1,0)$,
and
(ii)~$(|k_1|,|k_2|)=(2,0)$.

The case (ii) corresponds to $(\nu_1,\nu_2)=(\pm 1,\nu_2)$.
The corresponding Einstein metric (\ref{metric})
is independent of $\nu_2$ and
coincides with the standard $S^5$-metric
after a modification of the angle $\psi_1=2\tilde \psi_1$
($0\le\tilde\psi_1\le4\pi$):
\begin{eqnarray}
g&=&
d\theta^2
+\frac{1}{4}\sin^2\theta(d\tilde\psi_1+\cos\chi d\eta)^2
+\frac{1}{4}\cos^2\theta d\psi_2^2
+\frac{1}{4}\sin^2\theta g_{S^2}.
\label{metric:(ii)}
\end{eqnarray}
One can show that
the analysis remains true
even if the range is extended to $\nu_2>1$.

In the case (i), we have
$(\nu_1,\nu_2)=(\pm\frac{1}{2},0)$.
Then, the Einstein metric (\ref{metric})
is of the form
\begin{eqnarray}
g&=&
\frac{1-\frac{1}{4}\cos^2\theta}{1-\frac{1}{7}\cos^2\theta}d\theta^2
+\biggl(\frac{7}{16}\biggr)^2
 \frac{1-\frac{1}{7}\cos^2\theta}{1-\frac{1}{4}\cos^2\theta}
 \sin^2\theta(d\psi_1+\cos\chi d\eta)^2
\nonumber\\&&
+\frac{1}{4}\cos^2\theta d\psi_2^2
+\frac{7}{16}(1-\frac{1}{4}\cos^2\theta)g_{S^2},
\label{metric:(i)}
\end{eqnarray}
which gives a
metric of
cohomogeneity one
with principal orbits $ S^3\times S^1$.
The principal orbits
collapse to $ S^2\times S^1$ at $\theta=0$
and $S^3\times\{\mbox{point}\}$ at $\theta=\pi/2$.
Globally the metric can be regarded as one on
the non-trivial $S^3$-bundle over $S^2$.
In section 2, we will construct
Einstein metrics on $S^N$-bundles
over $S^2$ (see Theorem 3),
generalizing the metric (\ref{metric:(i)})
to higher dimensions.

\medskip

\top
\underline{\textbf{Case C}}

When we put $\nu_1=\nu_2\equiv\nu$,
the metric (\ref{metric})
is\fnote{$\sharp$}{
We have rescaled the metric as $\frac{2+\nu^2}{2}g\to g$.
}
\begin{eqnarray}
g_\nu=
d\theta^2
+\sum_{i,j=1,2}^2a_{ij}(\theta)\omega^i\otimes \omega^j
+\frac{2+\nu^2}{4(2\nu^2+1)}g_{S^2},
\label{g:nu:B}
\end{eqnarray}
where
\begin{eqnarray}
a_{11}
&=&
 \frac{\sin^2\theta}{4(2+\nu^2)}(2+\nu^2\cos^2\theta),\\
a_{22}
&=&
 \frac{\cos^2\theta}{4(2+\nu^2)}(2+\nu^2\sin^2\theta),\\
a_{12}
&=&
 -\frac{\nu^2}{4(2+\nu^2)}\sin^2\theta\cos^2\theta.
\end{eqnarray}
The 1-form $\omega^1\oplus\omega^2$ is a
connection
on the $T^2$-bundle
 over $S^2$,
locally written as
\begin{eqnarray}
\omega^i=d\psi_i+k\cos\chi d\eta~~~~~
(0\le\psi_i\le4\pi).
\end{eqnarray}
Here the coefficient $k$ is evaluated as
\begin{eqnarray}
k=\frac{\nu(\nu^2+2)}{2\nu^2+1},
\end{eqnarray}
by using (\ref{k}),
and it is required to be $k\in\bbZ$.
In this case
the $T^2$-bundles collapse to the same lens space $L_k$
at each boundary.
Notice that
by introducing the Maurer-Cartan forms of SU(2)
\begin{eqnarray}
\sigma_1&=&
2\cos
\left(\frac{\psi_1+\psi_2}{2}\right)d\theta
-\sin\left(\frac{\psi_1+\psi_2}{2}\right)
\sin\theta\cos\theta(d\psi_1-d\psi_2),\\
\sigma_2&=&
-2\sin\left(\frac{\psi_1+\psi_2}{2}\right)d\theta
-\cos\left(\frac{\psi_1+\psi_2}{2}\right)
\sin\theta\cos\theta(d\psi_1-d\psi_2),\\
\sigma_3&=&
\sin^2\theta d\psi_1+\cos^2\theta d\psi_2,
\end{eqnarray}
the fiber metric of (\ref{g:nu:B})
can be rewritten as
\begin{eqnarray}
g_F&=&
d\theta^2
+\sum_{i,j=1}^2 a_{ij}(\theta)d\psi_i\otimes d\psi_j
=\frac{1}{4}(\sigma_1^2+\sigma_2^2)+\frac{1}{2(2+\nu^2)}\sigma_3^2,
\end{eqnarray}
which reveals the SU(2) isometry of the metric.

\bigskip

\top
\textbf{Theorem 2.}~
\textit{
Let $\nu_k$ be real numbers satisfying
$k=\nu_k(\nu_k^2+2)/(2\nu_k^2+1)\in \bbZ$.
Then, $\{g_{\nu_k}\}$ gives an infinite series of
homogeneous
 Einstein metrics with positive scalar curvature 
$20(1+\nu_k^2)/(2+\nu_k^2)$
on $S^2\times S^3$.
}
\bigskip

\top
{\textit{proof.}}~~~
By the coordinate transformation,
\begin{eqnarray}
\alpha=2\theta,~~~
\beta=\frac{1}{2}(\psi_2-\psi_1)~~\mbox{and}~~t=\frac{1}{2}(\psi_1+\psi_2),
\end{eqnarray}
the metric takes the form
\begin{eqnarray}
g_{\nu_k}&=&
\frac{1}{4}(d\alpha^2+\sin^2\alpha d\beta^2)
+\frac{2+\nu^2}{4(2\nu^2+1)}(d\chi^2+\sin^2\chi d\eta^2)
\nonumber\\&&
+\frac{1}{2(2+\nu^2)}(dt+\cos\alpha d\beta +k\cos\chi d\eta)^2,
\end{eqnarray}
which represents a Kaluza-Klein metric on the total space $M^{1,1}_{k,1}$
of the
circle bundle
over $\CP1\times \CP1$
with Euler class $e=k\alpha_1+\alpha_2$,
where $\alpha_1$ and $\alpha_2$ are generators in
$H^2(\CP1 \times \CP1; \bbZ)=\bbZ\oplus\bbZ$. 
The space
$M^{1,1}_{k,1}$ is diffeomorphic to
$S^2\times S^3$ \cite{Wang}.
\hfill $\square$

\medskip
\top
\underline{\textit{Remark 4.}}~~
There exists the unique real number $\nu_k$ for each $k\in\bbZ$.
The value of $\nu_{k}$ is explicitly given by
\begin{eqnarray*}
\nu_k&=&
\frac{1}{6}[
{a^{1/3}}
-{8(3-2k^2)a^{-1/3}}
+{4k}
],
\\
a&=& -36k+64k^3+12\sqrt{96-183k^2+96k^4}
\end{eqnarray*}
for $k\ge 0$, and $\nu_{-k}=-\nu_k$.
\medskip

\top
\underline{\textit{Remark 5.}}~~
In the case $\nu_0=0$,
the metric coincides with  the product metric on $S^2\times S^3$:
\begin{eqnarray}
g_0=d\theta^2
+\frac{1}{4}\sin^2\theta d\psi_1^2
+\frac{1}{4}\cos^2\theta d\psi_2^2
+\frac{1}{2}g_{S^2}.
\end{eqnarray}

On the other hand,
in the limit $\nu_k\to\pm\infty$ ($k\to\pm\infty$),
the
fiber $S^1$ of $S^2\times S^3\to S^2\times S^2$ collapses,
and the  metric tends to the product Riemannian metric on $S^2\times S^2$
which is not Einstein:
\begin{eqnarray}
g_\infty=
d\theta^2
+\left(\frac{\sin\theta\cos\theta}{2}\right)^2 d\psi_-^2
+\frac{1}{8}g_{S^2}
\end{eqnarray}
with $\psi_-=\psi_1-\psi_2$.
\medskip

\sect{$d$-dimensional Einstein Metrics}
The AdS Kerr black hole in $d$-dimensions ($d\ge 4$)
was constructed in \cite{Hawking}.
It can be straightforwardly transformed to the Euclidean form,
\begin{eqnarray}
g_d&=&
 \frac{\Delta_r}{\rho^2}(
 d\tau
 -\frac{\alpha}{\Xi}\sin^2\theta d\phi
 )^2
+\frac{\Delta_\theta\sin^2\theta}{\rho^2}
 (
 \alpha d\tau
 +\frac{r^2-\alpha^2}{\Xi}d\phi
 )^2
\nonumber\\&&
+\frac{\rho^2}{\Delta_r}dr^2
+\frac{\rho^2}{\Delta_\theta}d\theta^2
+r^2\cos^2\theta g_{S^{d-4}},
\end{eqnarray}
where $g_{S^{d-4}}$ is
the standard metric
on $S^{d-4}$ with
the positive scalar curvature $(d-4)(d-5)$,
which is Einstein\fnote{$\flat $}{
The sphere metric $g_{S^{d-4}}$ can be replaced by $g_{M_{d-4}}$,
where $g_{M_{d-4}}$ is an arbitrary
Einstein metric on a $(d-4)$-dimensional manifold with
the positive scalar curvature $(d-4)(d-5)$.
},
and
\begin{eqnarray}
\rho^2&=&r^2-\alpha^2\cos^2\theta,\\
\Delta_r&=&(r^2-\alpha^2)(1-l^2r^2)-2Mr^{5-d},\\
\Delta_\theta&=&1-\alpha^2l^2\cos^2\theta
\end{eqnarray}
with the parameter $\Xi=1-\alpha^2l^2$.
We find that there exists a double root $r_0$
of $\Delta_r=0$,
when the following condition for the parameters
is satisfied:
\begin{eqnarray}
r_0&=&
 \left[
 \frac{d-3-(d-5)\nu^2}{d-1-(d-3)\nu^2}
 \right]^{1/2}l^{-1},\\
M_0&=&
 \frac{(1-\nu)^2}{d-1-(d-3)\nu^2}
 \left[
 \frac{d-3-(d-5)\nu^2}{d-1-(d-3)\nu^2}
 \right]^{\frac{d-3}{2}}l^{-(d-3)},\\
\Xi_0&=&
 1-\frac{\nu^2(d-3-(d-5)\nu^2)}{d-1-(d-3)\nu^2},
\end{eqnarray}
where we have introduced a dimensionless parameter $\nu=\alpha/r_0$.
Then, $\Delta_r$ takes the form
\begin{eqnarray}
\Delta_r&=&-(r-r_0)^2\tilde\Delta(r),\\
\tilde\Delta(r)&=&\frac{1}{r^{d-5}}(c_0+c_1r+\cdots+c_{d-3}r^{d-3}),
\label{Delta:tilde}
\end{eqnarray}
where
\begin{eqnarray}
c_i&=&
 \frac{2(1+i)(1-\nu^2)^2}{d-1-(d-3)\nu^2}r_0^{d-i-5}
 ~~~~~(0\le i\le d-6),\\
c_{d-5}&=&
 \frac{2d-8-3(d-5)\nu^2+(d-5)\nu^4}{d-1-(d-3)\nu^2},\\
c_{d-4}&=&
 2r_0l^2,\\
c_{d-3}&=&
 l^2.
\end{eqnarray}

The remaining procedure is completely parallel to the one
in the section 2;
consider a nearly extreme black hole,
and take the limit $\varepsilon\to 0$.
Actually we define new coordinates $(\eta,\chi,\phi_1)$
instead of $(\tau,r,\phi)$ as
\begin{eqnarray}
r&=&
 r_0-\varepsilon\cos\chi,\\
\eta&=&
 \frac{\varepsilon\tilde\Delta(r_0)}{r_0^2(1-\nu^2)}\tau
 +O(\varepsilon^2),\\
\phi_1&=&
 \phi+\frac{\alpha\Xi}{r_1^2-\alpha^2}
 ~~~~~~(r_1\equiv r_0-\varepsilon),
\end{eqnarray}
where $\tilde\Delta(r_0)$ can be calculated by (\ref{Delta:tilde});
\begin{eqnarray}
\tilde\Delta(r_0)&=&
 \frac{(d-5)(d-3)\nu^4-2(d-1)(d-5)\nu^2+(d-1)(d-3)}
 {d-1-(d-3)\nu^2}.
\label{Delta:tilde:0}
\end{eqnarray}

In the limit $\varepsilon\to 0$, we find a one-parameter
family of $d$-dimensional Einstein metrics
\begin{eqnarray}
g_\nu&=&
 h^2(\theta)d\theta^2
 +\sum_{i=1}^3a_i(\theta)\sigma^i\otimes\sigma^i
 +b^2(\theta)g_{S^{d-4}}.
\label{g:nu}
\end{eqnarray}
with the positive scalar curvature
${d(d-1)(d-3-(d-5)\nu^2)}/{(d-1-(d-3)\nu^2)}$.
Here $\sigma^i$ ($i=1,2,3$) are 1-forms defined by
\begin{eqnarray}
\sigma^1&=&\cos\psi d\chi +\sin\psi\sin\chi d\eta,\\
\sigma^2&=&-\sin\psi d\chi +\cos\psi\sin\chi d\eta,\\
\sigma^3&=&d\psi +k\cos\chi d\eta,
\end{eqnarray}
with $\phi_1=\frac{1}{2}(\psi+k\eta)$
and
\begin{eqnarray}
k&=&
 \frac{4\nu(d-1-(d-5)\nu^2)}{(d-5)(d-3)\nu^4-2(d-1)(d-5)\nu^2+(d-1)(d-3)}.
\label{k:nu}
\end{eqnarray}
The metric components are found to be
\begin{eqnarray}
h^2&=&
 \frac{1-\nu^2\cos^2\theta}{1-\mu^2\cos^2\theta},\\
a_1&=&
 a_2=\frac{1-\nu^2\cos^2\theta}{\tilde\Delta},\\
a_3&=&
 \frac{1}{4}
 \left(
 \frac{1-\nu^2}{1-\mu^2}
 \right)^2
 \frac{1-\mu^2\cos^2\theta}{1-\nu^2\cos^2\theta}
 \sin^2\theta,\\
b^2&=&\cos^2\theta,
\end{eqnarray}
where
\begin{eqnarray}
\mu^2&=&
 \frac{(d-3)\nu^2-(d-5)\nu^4}{d-1-(d-3)\nu^2},
\end{eqnarray}
and $\tilde \Delta$ is given by (\ref{Delta:tilde:0}).

To avoid singularities,
we will assume
$0\le\nu^2\le 1$.
In this range, $\mu^2$ and $k$
are monotonously
increasing functions with respect to $\nu$.
Then, we have $0\le\mu^2\le 1$,
 $-2\le k\le 2$
and $\tilde \Delta>0$ ($d\ge 4$).
By the analysis similar to section 2,
the ranges of angles must be restricted as $0\le\theta\le\frac{\pi}{2}$,
$0\le\psi\le 4\pi$, $0\le\chi\le\pi$ and
$0\le\eta\le 2\pi$.
If we impose that $k\in\bbZ$, the 1-form $\sigma^3$
can be regarded as a connection on the lens space $L_k$.
Taking account of the inequality $|k|\le 2$,
we have
\begin{eqnarray}
\mbox{(i)}~ |k|=0,~~~~~
\mbox{(ii)}~ |k|=1,~~~~~
\mbox{(iii)}~ |k|=2.
\nonumber
\end{eqnarray}

The case (i) corresponds to $\nu=0$,
and then the metric (\ref{g:nu})
gives the product metric on $S^2\times S^{d-2}$;
\begin{eqnarray}
g_0=
\frac{1}{d-3}
 \left(
 (\sigma^1)^2+(\sigma^2)^2
 \right)
+d\theta^2
+\frac{1}{4}\sin^2\theta d\psi^2
+\cos^2\theta g_{S^{d-4}}.
\end{eqnarray}

The case (iii) corresponds to $\nu=\pm 1$, and then
the metric (\ref{g:nu}) is the standard $S^d$-metric after a
modification of the angle $\psi\to 2\psi$;
\begin{eqnarray}
g_{\pm 1}&=&
d\theta^2
+\frac{1}{4}\sin^2\theta\left(
 (\sigma^1)^2+(\sigma^2)^2+(\sigma^3)^2
 \right)
+\cos^2\theta g_{S^{d-4}}.
\end{eqnarray}

In the case (ii),
the 1-forms $\sigma^i$ $(i=1,2,3)$
are identified
with the Maurer-Cartan forms of SU(2).
Thus the metric is of cohomogeneity one with
principal orbits $S^3\times S^{d-4}$.
The orbits collapse to $S^2\times S^{d-4}$ at $\theta=0$
and to $S^3\times\{\mbox{point}\}$ at $\theta=\pi/2$.
Hence, the total space is the unit sphere bundle
of the vector bundle $H\oplus\underline{\bbR}^{d-3}$
over $S^2$,
where $H$ is the Hopf bundle and $\underline{\bbR}^{d-3}$
is the trivial bundle of rank $d-3$.
Since the second Stiefel-Whitney class
$w_2(H\oplus\underline{\bbR}^{d-3})=w_2(H)\neq 0$
in $H^2(S^2;\bbZ)=\bbZ_2$,
the space is $S^2\widetilde{\times}S^{d-2} $,
the non-trivial $S^{d-2}$-bundle over $S^2$.

\bigskip

\top
\textbf{{Theorem 3}}~~~
\textit{Let $\nu$ be a real number satisfying
$\nu^2<1$ and
\begin{eqnarray}
\frac{4\nu(d-1-(d-5)\nu^2)}
{(d-5)(d-3)\nu^4-2(d-1)(d-5)\nu^2+(d-1)(d-3)}=\pm 1.
\end{eqnarray}
Then $g_\nu$ gives an Einstein metric with positive
scalar curvature on $S^{d-2}\widetilde{\times}  S^2$,
which  is the non-trivial $S^{d-2}$-bundle over $S^2$.
}
\bigskip

\top
\underline{\textit{Remark 6.}}~~~
For $d=4$, this result reproduces the Page metric on
$\CP2\sharp\overline{\CP2}$.
In this case
the principal orbits are $S^3$,
the range of  $\theta$ is extended to
$0\le\theta\le\pi$
and the metric has $\bbZ_2$-symmetry about $\theta=\pi/2$.
For $d=5$, this represents the metric (\ref{metric:(i)}).

\vspace{10mm}

\section*{Acknowledgements}
We thank R. Goto, H. Ishihara and S. Tanimura
for useful discussions.
Y.Y. would like to express his gratitude
to G.W. Gibbons and S.A. Hartnoll for
useful discussions
during his stay at DAMTP, Cambridge University.
This paper is supported by the 21 COE program
"Constitution of wide-angle mathematical basis focused on knots".
Research of Y.H. is supported  in part by the Grant-in
Aid for scientific Research (No.~15540090)
from Japan Ministry of Education.
Research of Y.Y. is supported  in part by the Grant-in
Aid for scientific Research (No.~14540073 and No.~14540275)
from Japan Ministry of Education.


\end{document}